\documentclass[a4paper,11pt]{article}
\pdfoutput=1 

\usepackage{jheppub} 

\usepackage[T1]{fontenc} 

\newcommand{\be}{\begin{equation}}
\newcommand{\ee}{\end{equation}}
\newcommand{\ba}{\begin{aligned}}
\newcommand{\ea}{\end{aligned}}

\def\cA{\mathcal{A}}
\def\cB{\mathcal{B}}

\def\cN{\mathcal{N}}
\def\cO{\mathcal{O}}

\def\cR{\mathcal{R}}
\def\cS{\mathcal{S}}

\DeclareMathOperator{\real}{Re}
\DeclareMathOperator{\im}{Im}

\DeclareMathOperator{\Li}{Li}

\def\Zpg{Z_\text{pert}^\text{geom}}
\def\Zpc{Z_\text{pert}^\text{conj}}
\def\geom{\text{geom}}
\def\conj{\text{conj}}
\def\pert{\text{pert}}

\title{\boldmath S-duality resurgence in SL(2) Chern-Simons theory}

\author[a]{Dongmin Gang}  
\author[b]{and Yasuyuki Hatsuda}

\affiliation[a]{Center for Theorectical Physics, Seoul National University \\Seoul 08826, Korea}
\affiliation[b]{Department of Physics, Rikkyo University,\\Toshima, Tokyo 171-8501, Japan}

\emailAdd{arima275@snu.ac.kr}
\emailAdd{yhatsuda@rikkyo.ac.jp}

\preprint{RUP-17-22}

\abstract{We find that an $S$-duality in $SL(2)$ Chern-Simons theory for hyperbolic 3-manifolds emerges by the Borel resummation of a semiclassical expansion around a particular flat connection associated to the hyperbolic structure. We demonstrate it numerically with two representative examples  of hyperbolic 3-manifolds.  }

\begin{document} 
\maketitle
\flushbottom

\section{Introduction and Summary}
Perturbative expansions in quantum mechanics/quantum field theories are in general asymptotic expansions with zero radius of convergence. 
Typically, their coefficients grow factorially. To know the information on physical observables at finite coupling, we thus need a resummation method of asymptotic expansions. 
A systematic approach to construct a complete trans-series expansion in general situation is a so-called {\it resurgent analysis} (For reviews, see \cite{Marino:2012zq, Dorigoni:2014hea, Dunne:2015eaa} for instance). 
The resurgence theory implies that the perturbative sector and the non-perturbative sectors are not independent but interrelated to each other.

The Chern-Simons (CS) theory is an example of exactly solvable quantum field theories \cite{witten1989quantum} and its perturbative/non-perturbative aspects  have been extensively studied last three decades in various contexts of  theoretical physics. Applications of the topological theory include 3 dimensional quantum gravity \cite{Witten:1988hc}, topological strings \cite{Witten:1992fb}, 3 dimensional superconformal field theories \cite{Terashima:2011qi,Dimofte:2011ju} and mathematical physics \cite{1996q.alg.....1025K}.

It is a natural idea to apply  the resurgence technique to the Chern-Simons theory and see how the resurgence  helps us to understand (or find) some aspects (new aspects) of  Chern-Simons theory. From this motivation, a refinement of a CS invariant was addressed in \cite{Gukov:2016njj} (see also \cite{Honda:2016vmv, Honda:2017qdb} for a different perspective). In this paper, we study another mysterious aspect of CS theory, an $S$-duality  \cite{Dimofte:2011jd} when the gauge group is complex $SL(2)$. Although there are already several hints on the $S$-duality from state-integral models for the complex CS theory, 3d/3d correspondence and etc,  our resurgent analysis gives more direct evidence and more precise statement for it. 
Note that the similar hidden $S$-duality structure also appears in the context of the so-called ``Topological Strings/Spectral Theory'' correspondence \cite{Grassi:2014zfa, Wang:2015wdy, Hatsuda:2015qzx, Franco:2015rnr}.

Let us briefly summarize our main statement in this paper.
We find that the perturbative expansion around a saddle point corresponding to a particular flat connection, $A=A^{\rm conj}$ defined in \eqref{Agoem/conj}, is \textit{Borel summable}, and its Borel resummation has the $S$-duality property, while the resummations around the other saddle points do not. State-integral models do not seem to provide its simple explanation.%
\footnote{Integrands in state-integral models have such an $S$-daul symmetry manifestly, but a choice of integration contours may break it.} 
In 3d/3d correspondence \cite{Terashima:2011qi,Dimofte:2011ju}, the $S$-duality is related to the manifest $b\leftrightarrow 1/b$ symmetry  of a curved background called  squashed 3-sphere  $S^3_b$ \cite{Hama:2011ea} where $b$ denotes a squashing parameter. Our analysis also provides supporting evidence  for the conjecture in \cite{Andersen:2011bt,Gang:2014ema,Bae:2016jpi} (see also recent discussion in \cite{Mikhaylov:2017ngi}) saying that only the flat connection $A^{\rm conj}$ on $M$ contributes to the $S^3_b$ partition function of the corresponding 3d theory $T[M]$ in \eqref{T[M]}. 

The rest of the paper is organized as follows. In section \ref{sec : perturbation}, we introduce two $SL(2)$ flat connections, $A^{\rm geom}$ and $A^{\rm conj}$, on hyperbolic 3-manifolds $M$ and perturbative expansions around them. As tools to compute the perturbative expansions, the volume conjecture and state-integral models are reviewed. In section \ref{sec : resummation}, we perform the Borel-Pad\'e resummation of  the perturbative expansions for two hyperbolic 3-manifolds, the figure-eight knot complement and a closed 3-manifold called {\it Thurston} manifold,  and check the $S$-duality of the resummation of the perturbative expansion around $A^{\rm conj}$. We also provide a heuristic understanding of the $S$-duality by embedding the perturbative expansion to an unitary complex CS theory where the symmetry is manifest in Lagrangian. 

\section{SL(2) Chern-Simons perturbation series on hyperbolic 3-manifolds} \label{sec : perturbation}
\subsection{Perturbative invariants from Complex Chern-Simons theory}
We consider an  asymptotic expansion of following {\it formal} path integral 
\begin{align}
Z_{\rm pert}(k;M) = \int [DA] e^{\frac{i k}{4\pi} CS[A;M]}\;. \label{Formal path-integral}
\end{align}
Here $A$ is an $SL(2)$ gauge field on  a 3-manifold $M$ and $S[A;M]$ is the Chern-Simons functional
\begin{align}
CS[A;M]:= \int_M \textrm{Tr}(dA+\frac{2}3 A^3)\;.
\end{align}
In perturbation, we need to choose a flat-connection $A^\alpha$ and let the formal perturbative expansion in $1/k$ around it be
\begin{align}
Z_{\rm pert}^{\alpha} (k)\;.
\end{align}
For a hyperbolic 3-manifold $M$, there are two special $SL(2,\mathbb{C})$ flat connections, $A^{\rm geom}$ and $A^{\rm conj}$,  associated to the unique hyperbolic metric on $M$ normalized as $R_{\mu\nu}= - 2g_{\mu\nu}$.
\begin{align}
A^{\rm geom} = \omega + i e\;, \quad A^{\rm conj} = \omega- i e\;. \label{Agoem/conj}
\end{align}
Here $\omega$ and $e$ are a spin-connection and a dreibein respectively constructed from the hyperbolic metric. Both of them can be considered as $so(3)$-valued 1-forms and they form an $sl(2)$-valued 1-form. The hyperbolicity condition, $R_{\mu \nu}= -2 g_{\mu \nu}$, implies that both of $A^{\rm geom}$ and $A^{\rm conj}$ are flat connections. One basic characteristic of them is that $A^{\rm geom}$ ($A^{\rm conj}$) gives the exponentially largest (smallest) classical contribution for real large $k \in \mathbb{R}_+$:
\begin{align}
\im \big{(} CS[A^{\rm geom};M] \big{)}\leq \im CS[A^{\alpha};M] 
\leq \im \big{(} CS[A^{\rm conj};M] \big{)}
\label{eq:ineq-CS}
\end{align}
for any flat connection $A^\alpha$. In particular, we have
\begin{align}
\im \big{(} CS[A^{\rm conj};M] \big{)} =  -\im \big{(} CS[A^{\rm geom};M] \big{)} = 2 \textrm{vol}(M)\;.
\label{range of ImCS}
\end{align}
Here $\textrm{vol}(M)$ is a topological invariant called {\it hyperbolic volume} defined as the volume measured in the unique hyperbolic metric. 
For these isolated irreducible flat connections, the perturbative expansion takes the following form \cite{Gukov:2016njj}
\begin{align}
Z_{\rm pert}^{\alpha} (k;M) = e^{\frac{i k}{4\pi} CS[A^\alpha;M]} \sum_{n \geq 0}\frac{a^{\alpha}_n}{k^n}\;, \quad 
\;.
\label{eq:pert}
\end{align}
The two perturbative expansions for $\alpha$ = geom or conj are especially related by
\begin{align}
a_n^{\rm conj} = (-1)^n (a_n^{\rm geom})^*\;.
\label{eq:geom-conj}
\end{align}
In principle, the formal perturbative expansion around a given flat connection can be computed by summing up contributions from Feynman diagrams. For the computation, we need to fix a gauge symmetry by introducing a metric on a 3-manifold. The final sum should be independent on the  choice due to the topological property of the theory but each contribution  might depend on the  choice and the computation requires the full knowledge on the spectrum of Laplacian on the 3-manifold with respect to the metric as for usual quantum field theories. There are simpler methods fully using topological property  of the Chern-Simons theory. In the subsequent sections, we review two approaches. 

\subsection{From $su(2)$ knot/3-manifold invariants}
One simple way of computing the perturbative invariant  is to use an asymptotic limit of $su(2)$ knot/3-manifold invariants called colored Jones polynomial/Witten-Turaev-Rashetikin invariants (WRT) \cite{JonesPolynomial,Witten:1988hf,ReshetikhinTuraev,KirbyMelvin} assuming the volume conjecture \cite{1996q.alg.....1025K,1999math......5075M,2008LMaPh..86...79G,ChenYang,Gang:2017cwq}. 
\paragraph{Volume conjecture.}  In an asymptotic limit $N \in \mathbb{Z}\rightarrow \infty$,
\begin{align}
&\frac{1}{N^{3/2}}\frac{J_N \big{(}q=\exp(\frac{2\pi i}{N});K \big{)}}{J_N \big{(}q=\exp(\frac{2\pi i}{N});({\rm unknot}) \big{)}} \sim Z_{\rm pert}^{\rm geom}(k=N;M=S^3\backslash K)\;,
\end{align}
for a hyperbolic knot $K$ in $S^3$. Similarly, in an asymptotic limit $N\in 2 \mathbb{Z}+1 \rightarrow \infty$,
\begin{align}
&\tau^{\rm SO(3)}_N (M)\sim Z_{\rm pert}^{\rm geom}(k=N;M)\;,
\end{align}
for a hyperbolic closed 3-manifold $M$. 
Here $\sim$ means both sides have the same asymptotic expansion in $1/N$, 
and $J_N(q;K)$ is a quantum knot invariant called  {\rm colored Jones polynomial} of a knot $K$ and $\tau_{N}^{SO(3)}(M)$ is an $SO(3)$ version of WRT invariant. 
For example, 
\begin{align}
\begin{split}
&J_N (q;K=\textrm{unknot}) = \frac{q^{N/2}-q^{-N/2}}{q^{1/2}-q^{-1/2}}\;,
\\
&J_N (q;K=\mathbf{4_1}) = \frac{q^{N/2}-q^{-N/2}}{q^{1/2}-q^{-1/2}} \sum_{j=0}^{N-1}\prod_{i=1}^j \big{(}q^{(N-i)/2}-q^{-(N-i)/2}\big{)} \big{(}q^{(N+i)/2}-q^{-(N+i)/2}\big{)}\;.
\end{split}
\end{align}
Here $\mathbf{4_1}$ denotes the figure-eight knot, the simplest hyperbolic knot. For a closed 3-manifold $M=(S^3 \backslash  K)_p$ obtained by taking Dehn surgery along a knot $K$ with slope $p\in \mathbb{Z}$, the $SO(3)$ WRT invariant is given as following formula
\begin{align}
\tau^{SO(3)}_N (M) = \frac{2}N e^{\pi i \big{(} \frac{3+r^2}r+ \frac{r-3}4 \big{)}} \bigg{(} \sum_{r=1}^{N-1} \sin^2 ( \frac{2r\pi}{N}) (-e^{\frac{\pi i}{N}})^{-p (N^2-1)} J_N (q= e^{\frac{2\pi i}N};K)\bigg{)}\;.
\end{align}

\subsection{From $SL(2)$ state-integral models}
Another simple approach is to use state-integral models based on ideal triangulation and Dehn filling representation of 3-manifolds. Decomposing a 3-manifold into basic building blocks, ideal tetrahedra and solid-torus, the $SL(2)$ CS partition functin can be computed by {gluing the wave-functions on them. As  a topological field theory, the phase spaces associated to the boundaries of basic building blocks are finite dimensional non-compact symplectic varieties and the wave functions depend on the finite number of continuous position variables and the gluing of the wave functions is  realized as an integration over the boundary variables. We refer to \cite{2007JGP,Dimofte:2011gm,Andersen:2011bt,Dimofte:2012qj,Dimofte:2014zga} for state-integral models for knot complements based on its ideal triangulation and its extension \cite{Bae:2016jpi,Gang:2017cwq}  to closed 3-manifolds by incorporating Dehn filling operation. We  give  explicit expressions for the state-integral model for two simple hyperbolic 3-manifolds, the figure-eight knot complement and a closed hyperbolic 3-manifold called  {\it Thurston} manifold.  

\paragraph{Figure-eight knot complement.} For $M=\textrm{(figure-eight knot complement)}= S^3\backslash \mathbf{4_1}$, the state-integral model is given by \cite{Dimofte:2011gm}
\begin{align}
Z(k;M=S^3\backslash \mathbf{4_1})(u) = e^{ - \frac{u^2+(2i \pi +\hbar) u}{2\hbar} }   \int \frac{\sqrt{k}dz}{2\pi} \frac{\Psi_\hbar (z-u)}{\Psi_\hbar(-z)} e^{\frac{z u}{\hbar}}\;, \label{state-integral-41}
\end{align}
where
\begin{align}
\hbar := \frac{2\pi i}k \;.
\end{align}
The 3-manifold has a torus boundary and there is a conventional canonical choice for basis of the boundary 1-cycles called {\it meridian} ($\mu$) and {\it longitude} ($\lambda$). %
\begin{align}
\partial M = \mathbb{T}^2\;, \quad H_1 (\partial M, \mathbb{Z}) = \mathbb{Z}\times \mathbb{Z} = \{  p \mu+ q\lambda : p, q\in \mathbb{Z} \}\;,
\end{align}
and the $u$ parametrizes the  fixed boundary  $SL(2)$ holonomy around the meridian cycle
\begin{align}
 P \exp \oint_{\rm merdian} A \sim \left( {\begin{array}{cc}
   e^{u/2} & 1 \\
   0 & e^{-u/2} \\
  \end{array} } \right)\;,
\end{align}
and the state-integral is invariant under the  Weyl-symmetry, $u\leftrightarrow -u$. 
Here $\sim$ denote the equivalence relation by $SL(2)$ conjugation. 
The function $\Psi_{\hbar}(z)$ is related to the non-compact quantum dilogarithm $\Phi_b(z)$, defined in \eqref{eq:NQDL}, by
%
\begin{align}
\Psi_{\hbar}(z):=\Phi_b\left( \frac{z}{2\pi b} \right), \qquad \hbar=\frac{2\pi i}{k} = 2\pi i b^2  .
\end{align}
%
This function has the following  interesting $S$-duality 
\begin{align}
\Psi_\hbar (z) = \Psi_{-\frac{4\pi^2}\hbar} (\frac{2\pi i z}\hbar)\;.
\end{align}
We will discuss some basic properties of the quantum dilogarithm in appendix~\ref{app:QDL}.
Using the semiclassical expansion of $\Phi_b(z)$ (see \eqref{eq:sc-1}), we have
\begin{align}
\log \Psi_\hbar (z)  = \sum_{n\geq 0}\hbar^{n-1} \frac{B_{n}(1/2)}{n!} \Li_{2-n}(-e^{z}),\qquad \hbar \to 0.
\end{align}
where $B_n(x)$ is the $n$-th Bernoulli polynomial.
Note that $B_n(1/2)$ is vanishing for all odd $n$.
The state-integral is then written in the following form
\begin{align}
Z\big{(}k;M=S^3\backslash \mathbf{4_1}\big{)}(u) = \int \frac{\sqrt{k}dz}{2\pi} \exp \bigg{(} \sum_{n \geq 0} \hbar^{n-1} W_{n} (z,u;S^3\backslash \mathbf{4_1})\bigg{)}\;, \label{Perturbative state-integral}
\end{align}
where the leading classical part is 
\begin{align}
W_0(z,u;S^3\backslash \mathbf{4_1}) = - \frac{1}2 u^2 + i \pi u + z u + \Li_2 (-e^{z-u})-\Li_2 (-e^{-z})\;.
\end{align}
At $u=0$, there are two saddle points, $z^{\rm geom}$ and $z^{\rm conj}$
\begin{align}
z^{\rm geom} = -\frac{2 \pi i} 3\;, \quad z^{\rm conj} = \frac{2 \pi i} 3\;. \label{two saddles in 41}
\end{align} 
Expanding the integrand in \eqref{Perturbative state-integral} around these saddle points, one obtains
the following perturbative expansions:
\begin{align}
\begin{split}
Z^{\rm geom}_{\rm  pert} (k;S^3\backslash \mathbf{4_1})(u=0)
&= \frac{e^{\frac{kV}{2\pi}}}{3^{1/4}} \bigg{(}1+ \frac{11\pi}{36 \sqrt{3}k }+\frac{697\pi^2}{7776 k^2}+\ldots \bigg{)}\;,
\\
Z^{\rm conj}_{\rm  pert} (k;S^3\backslash \mathbf{4_1})(u=0)
&= \frac{e^{-\frac{kV}{2\pi}}}{3^{1/4}} \bigg{(}1- \frac{11\pi}{36 \sqrt{3}k }+\frac{697\pi^2}{7776 k^2}-\ldots \bigg{)}\;.
\end{split}
\label{eq:pert-fig8}
\end{align}
where $V$ is the hyperbolic volume of the knot complement:
\begin{align}
V=\textrm{vol}(S^3\backslash \mathbf{4_1}) = 2 \im [\Li_2 (e^{\pi i/3})] = 2.02988\ldots \;.
\end{align}
The state-integral can be interpleted as  a squashed 3-sphere partition function of a 3d $\mathcal{N}=2$ gauge theory $T[S^3\backslash \mathbf{4_1}]$ associated to the knot complement upon a proper choice of integral contour. In this identification, the  formal $SL(2)$ CS level $k$ is related to the squashing parameter $b$ by the relation $k=b^{-2}$.
Following \cite{Dimofte:2011ju}, we define 
\begin{align}
\begin{split}
T[M]:=& \textrm{3d  theory obtained from a twisted compacitification }
\\
&\textrm{of 6d $A_1$ (2,0) theory on a hyperbolic 3-manifold  $M$\;.} \label{T[M]}
\end{split}
\end{align}
According to  \cite{Dimofte:2011ju}, 
\begin{align}
T[S^3\backslash \mathbf{4_1}]= (\textrm{$u(1)_0$ gauge theory coupled to two chrial multiplets of charge $+1$} )\;.
\end{align}
The subscript in $u(1)_0$ denotes the CS level for the gauge $u(1)$ symmetry. The theory has  $SU(2)_\Phi \times U(1)_J$ symmetry where the $SU(2)$ rotates two chiral fields and $U(1)_J$ is  the topological symmetry whose conserved charge is the monopole flux of the gauge $U(1)$. It is argued  that the symmetry is enhanced to $SU(3)$ at the IR fixed point  \cite{ToAppear}. 
To find a  contour of the state-integral relevant to the gauge theory, let us first briefly summarize the localization on $S^3_b$ \cite{Hama:2011ea} in our notation.
\begin{align}
\begin{split}
&\bullet \textrm{ a free chiral $\Phi$ of R-charge $\Delta$ and charge $q$ under a $u(1)$ symmetry} \;:
\\ 
&\quad \quad \textrm{$\exp \left( - \frac{i\pi}2 \big{(}q \sigma - \frac{i (b+b^{-1}) }2 (1-\Delta)  \big{)}^2 \right)  \Psi_{\hbar} \bigg{(}-2\pi  q b   \sigma + (\pi i +\frac{\hbar}2 )(1-\Delta) \bigg{)}$}\;,
\\
&\bullet \textrm{ gauging the $u(1)$}  \;:\; \int_{\mathbb{R}} d \sigma\;,
\\
&\bullet \textrm{ CS term for the $u(1)$ with level $k$}  \;:\; \exp (-i \pi k \sigma^2)\;.
\\
&\bullet \textrm{ Fayet-Iliopoulos (FI) term  for the $u(1)$ with  parameter $\zeta$} \; :\; \exp (- 2i \pi \zeta \sigma)
\end{split}
\end{align}
Here $\sigma$ is a real scalar in a vector multiplet coupled to the $u(1)$ symmetry. Applying   the localization formulae, 
\begin{align}
\begin{split}
&(\textrm{partition function for $T[S^3 \backslash \mathbf{4_1}]$ on $S^3_b$}) 
\\
&= \int_{\mathbb{R}} d\sigma \Psi_\hbar \left(-2\pi b (\sigma + \frac{\zeta_1}2 )+ (\pi i +\frac{\hbar}2) (1-\Delta_1) \right) \Psi_\hbar \left(-2\pi b (\sigma -\frac{\zeta_1}2 )+ (\pi i +\frac{\hbar}2)(1-\Delta_2) \right) 
\\
&\quad \times   \exp \left(- \frac{i \pi}2 (\sigma +\frac{\zeta_1}2 - i (b+b^{-1}) (1-\Delta_1)^2)  - \frac{i \pi}2 (\sigma- \frac{\zeta_1}2 - i (b+b^{-1}) (1-\Delta_2)^2)\right)
\\
&\quad \times \exp \big{(} -2 i \pi  \zeta_2 \sigma  \big{)}\;.
\end{split}
\end{align}
Here $\Delta_1$ and $\Delta_2$ are the R-charge choices for two chiral multiplets. $\zeta_1$ is the real mass for  a Cartan $u(1)$ of the $SU(2)$ flavor symmetry and  the $\zeta_2$ is the FI parameter, which can be considered as the real mass for the  $u(1)_{J}$ symmetry. This expression is equivalent to the state-integral \eqref{state-integral-41} when we choose 
\begin{align}
\Delta_1 = \Delta_2  = \frac{1}3\;, \quad \zeta_2 = \frac{3}2 \zeta_1\;,
\end{align}
with the following change of variables
\begin{align}
\zeta_1 = -\frac{u}{2\pi b}\;, \quad z = - 2\pi b (\sigma+ \frac{\zeta_1}2)+ \frac{2\pi i}3 (1+b^2)\;. \label{specialization}
\end{align}
Since the $\sigma$ and $\zeta_i$ are real variables, the relation tell us  that the state-integral model can be  interpreted as  $S^3_b$ partition function of $T[S^3\backslash \mathbf{4_1}]$ when integrated over following contour
\begin{align}
\Gamma_{S^3 \backslash \mathbf{4_1}} =\mathbb{R}+ \frac{2\pi i }3  (1+b^2)\;. \label{Gamma41}
\end{align}
Note that the saddle point $z^{\rm conj}$ in \eqref{two saddles in 41} asymptotically touch the contour in the  limit $b\rightarrow 0$. This may imply that the squashed 3-sphere partition function of $T[S^3\backslash   \mathbf{4_1}]$ is  asymptotically equal to $Z^{\rm conj}_{\rm pert}[S^3\backslash \mathbf{4_1}]$ in the $b\rightarrow 0$ limit. 
 In this case, as a side remark,  the {\it geometric} R-charge choice\footnote{{\it geometric} R-charge choice means a R-charge choice under which the $S^3_b$ partition function of the gauge theory $T[M]$ can be made to to be identical to the $SL(2)$ CS  state-integral model $Z(k;M)$ upon a proper choice of integration contour. }  coincides with  the {\it conformal} R-charge at infrared (IR) fixed point and the state-integral at $u=0$  gives the $S^3_b$ partition function of the IR superconformal field theory.

\paragraph{Thurston manifold.} Let $(S^3\backslash \mathbf{4_1})_p$ be a closed 3-manifold obtained by Dehn surgery along figure-eight knot with integral slope $p$. 
\begin{align}
\begin{split}
&(S^3 \backslash \mathbf{4_1})_p :=\big{ [}(S^3 \backslash \mathbf{4_1}) \bigcup (D_2 \times S^1) \big{]}/\sim\;,  \textrm{ where we identify }
\\
& \big{(} p \mu +\lambda \in H_1 (\partial (S^3 \backslash \mathbf{4_1}) ,\mathbb{Z})  \big{)}\sim  \big{(} \textrm{contractible boundary $S^1$} \subset \partial (D_2 \times S^1) \big{)}
\end{split}
\end{align}
The  state-integral model for the closed 3-manifold is \cite{Bae:2016jpi}
\begin{align}
\begin{split}
&Z\big{(}k;M=(S^3\backslash \mathbf{4_1})_p \big{)} = \int \frac{\sqrt{k}du}{\pi} \exp(\frac{p u^2}{4\hbar}) \sinh(\frac{u}2) \sinh(\frac{\pi i u}\hbar)  Z(k;M=S^3\backslash \mathbf{4_1})(u) \;,
\\
&= \int \frac{k dz du }{2\pi^2} \exp \left(\frac{(p-2) u^2-(4\pi i +2 \hbar) u + 4 z u}{4\hbar}\right) \sinh(\frac{u}2) \sinh(\frac{\pi i u}\hbar)    \frac{\Psi_\hbar (z-u)}{\Psi_\hbar(-z)} \;.  \label{state-integral-thurston1}
\end{split}
\end{align}
When $p=-5$, the 3-manifold $(S^3\backslash \mathbf{4_1})_{p=-5}$ is called the Thurston manifold, which is known to be the second smallest hyperbolic 3-manifold with 
\begin{align}
{\rm vol}\big{(}(S^3\backslash \mathbf{4_1})_{p=-5} \big{)}=0.981369\ldots\;.
\end{align}
In this case, there are two saddle points $(z^{\rm conj_\pm},u^{\rm conj_\pm})$ corresponding to the flat connection $A^{\rm conj}$ 
\begin{align}
\begin{split}
&(z^{{\rm conj}_+}, u^{{\rm conj}_+}) = (-0.929172 + 1.90501 i, -0.721568 - 1.15121 i)\;,
\\
&(z^{{\rm conj}_-}, u^{{\rm conj}_-}) = (1.59632 + 2.79266 i, 0.721568 + 1.15121 i)\;.
\end{split}
\end{align}
Two saddle points are related by the Weyl-reflection  of $SL(2)$ and  the perturbative expansions around two saddle points are identical to all order 
\begin{align}
\begin{split}
&Z^{{\rm conj}_+}_{\rm pert} \left(k;(S^3\backslash \mathbf{4_1})_{p=-5} \right) = Z^{{\rm conj}_-}_{\rm pert} \left(k;(S^3\backslash \mathbf{4_1})_{p=-5} \right)
\\
&=(-0.0512672 - 0.350846 i) \exp \left( \frac{k(- 0.981369+1.52067i) }{2\pi }\right) \\
&\qquad \times
\left( 1+\frac{-0.0975308 + 0.0782969 i}{k}+\frac{-0.364363 + 0.236171 i}{k^2}+\ldots \right) \;.
\end{split}
\label{eq:pert-Thurston}
\end{align}
%
\\
The  $Z^{\rm conj}_{\rm pert}\left(k;(S^3\backslash \mathbf{4_1})_{p=-5} \right)$ is the sum of contributions from two saddle points
\begin{align}
\begin{split}
Z^{{\rm conj}}_{\rm pert} \left(k;(S^3\backslash \mathbf{4_1})_{p=-5} \right) &= Z^{{\rm conj}_+}_{\rm pert} \left(k;(S^3\backslash \mathbf{4_1})_{p=-5} \right)+Z^{{\rm conj}_-}_{\rm pert} \left(k;(S^3\backslash \mathbf{4_1})_{p=-5} \right)
\\
&=2Z^{{\rm conj}_+}_{\rm pert} \left(k;(S^3\backslash \mathbf{4_1})_{p=-5} \right)
\end{split}
\end{align}
Interestingly, there is a simper integral expression which reproduce the same perturbative expansion \cite{Gang:2017lsr}. Let
\begin{align}
\widetilde{Z}\left(k;(S^3\backslash \mathbf{4_1})_{p=-5}\right) = \sqrt{\frac{i k}{2\pi^2}} \int dz \exp \left(-\frac{(2\pi i+ \hbar)^2}{8\hbar}+ \frac{(2\pi i +\hbar) z}\hbar- \frac{3 z^2}{2\hbar} \right) \Psi_\hbar (z)\;. \label{state-integral-thurston2}
\end{align}
%
One saddle point for the integral is
\begin{align}
z^{\rm conj}=-0.061412 + 1.8063142i\;. \label{eq:z-saddle-conj}
\end{align}
One can check that the perturbative expansion of $\widetilde{Z}$ around the saddle point gives the same perturbative expansion with $Z^{{\rm conj}}_{\rm pert} \left(k;(S^3\backslash \mathbf{4_1})_{p=-5} \right)$. With a proper choice of integral contour, the state-integral $\widetilde{Z}$ can be interpreted as the partition function of a 3d gauge theory $T[{\it Thurston}]$ on a squashed 3-sphere.    The theory $T[{\it Thurston}]$ is field-theorectically described as \cite{Gang:2017lsr}
\begin{align}
T[{\it Thurston}] = u(1)_{-7/2} \textrm{ coupled to a chrial $\Phi$}\;.
\end{align}
From a localization, we have
\begin{align}
\begin{split}
&(\textrm{partition function for $T[{\it Thurston}]$ on $S^3_b$}) 
\\
&= \int_{\mathbb{R}} d \sigma   \exp \left( - \frac{i\pi}2 \big{(} \sigma - \frac{i (b+b^{-1}) }2 (1-\Delta)  \big{)}^2+\frac{7}2 \pi i \sigma^2 \right) \Psi_\hbar \bigg{(}-2\pi b \sigma+ (\pi i+\frac{\hbar}2 )(1-\Delta) \bigg{)}
\end{split}
\end{align}
Replacing the integration variable $\sigma$ by $z:=-2\pi b \sigma+(\pi i +\frac{\hbar}2)(1-\Delta)$, the integral become
\begin{align}
\int_{\mathbb{R}+ \pi i (1+ b^2) (1-\Delta)} \frac{d z}{(2\pi b)} \exp \left(  -\frac{7 (1-\Delta)^2 (2\pi i  + \hbar)^2}{16 \hbar} + \frac{7(1-\Delta) (2\pi i +\hbar)z}{4\hbar} - \frac{3z^2}{2\hbar} \right)  \Psi_\hbar (z)
\end{align}
Choosing $\Delta =\frac{3}7$, the integral become identical to the state-integral $\widetilde{Z}$ modulo following factor
\begin{align}
\sqrt{2} \times \textrm{exp}\left( i \pi (b^2+b^{-2}) Q_1 + i \pi Q_2\right)\;, \quad Q_1,Q_2 \in \mathbb{Q}\;.
\end{align}
Except the factor $\sqrt{2}$, the remaining factor is purely phase factor  which can be removed by  a local counterterm  and  thus negligible.  The factor $\sqrt{2}$ may come from a topological degree of freedom coupled to the system. Modulo the contribution from topological degree of freedom, the $\widetilde{Z}$ is $S^3_b$-partition function of the $T[{\it Thurston}]$ theory when integrated over following contour
\begin{align}
\Gamma_{\rm Thurston} =\mathbb{R}+ \frac{4\pi i }7  (1+b^2)\;. \label{Gamma-thurston}
\end{align}
Note that $\frac{4\pi}7=1.7952..$. So the contour $\Gamma_{\rm Thurston}$ is very close to the saddle point in  \eqref{eq:z-saddle-conj}  in the limit $b\rightarrow 0$ and  can be smoothly deformed to touch the saddle point.
In this case, the {\it geometric} R-charge choice ($\Delta=\frac{3}7$) is different from the IR conformal R-charge determined by F-maximization \cite{Jafferis:2010un}. 

\section{Resumming perturbative CS invariants and $S$-duality} \label{sec : resummation}
\subsection{Borel resummation method}  
Here we discuss the resummation for the perturbative expansions \eqref{eq:pert-fig8} and \eqref{eq:pert-Thurston} 
(or \eqref{eq:pert} more generally). 
The important fact is that all of these perturbative expansions are \textit{divergent} series.
Therefore one needs a resummation method to get a finite value for given $k$.
The standard way to do so is the Borel summation method.
We briefly review it at the beginning in this section.

Let us consider a formal perturbative series of the form
\be
f(k)=\sum_{n=0}^\infty \frac{f_n}{k^n},\qquad  k \to \infty .
\label{eq:f-pert}
\ee
We assume that the perturbative coefficient $f_n$ factorially diverges in $n \to \infty$. 
Therefore this perturbative expansion is a formal divergent series.
The Borel transform of this series expansion is defined by
\be
\cB f (\zeta):=\sum_{n=0}^\infty \frac{f_n}{n!}\zeta^n \:.
\label{eq:Borel-trans}
\ee
Note that this infinite sum is now \textit{convergent}.
We can analytically continue it to the complex $\zeta$-plane except for its singularities.
We then define the Borel sum by the Laplace transform:
\be
\cS f (k) := k \int_0^\infty d\zeta\,  e^{-k\zeta} \cB f(\zeta)\:.
\ee
The asymptotic expansion of this Borel sum reproduces the original divergent series \eqref{eq:f-pert}.
The Borel sum gives a meaning of the formal divergent series.
If there are no singularities on the integration contour (i.e., on the positive real axis),
the Laplace transform in the Borel sum is well-defined. In this case, $f(k)$ is called Borel summable.
However, we often encounter the situation that the integrand has singularities on $\zeta \in \mathbb{R}_+$. 
This case is called non Borel summable.
In the non Borel summable case, we deform the integration contour,
and define a new deformed Borel sum by
\be
\cS_\theta f (k) :=k \int_0^{\infty e^{i\theta}} d\zeta\,  e^{-k\zeta} \cB f(\zeta)\:,
\ee
where $\theta$ is chosen to avoid the singularities.
In our case, it is sufficient to consider the case where $\theta$ is very close to $0$
in order to avoid singularities on the positive real axis.
We denote it as
\be
\cS_{\pm} f(k)= \cS_{\theta=\pm \epsilon} f(k), \qquad \epsilon>0 \:,
\label{eq:lateral-Borel}
\ee
where $\epsilon$ is a small constant.
Unless the contour hits a singularity, the Laplace intergal does not  
depend on $\epsilon$.
If the Borel transform has singularities on the positive real axis,
the deformed Borel sums $\cS_\pm f(k)$ do not agree with each other:
\be
\cS_+ f(k) \ne \cS_- f(k) \:.
\ee
The discontinuity of the Borel sums is called the Stokes phenomenon.

In practical computations, we know only the first several values of $f_n$.
If we have $f_n$ up to $n=2n_\text{max}$,
then the Borel transform \eqref{eq:Borel-trans} is truncated at $n=2n_\text{max}$:
\be
\cB f(\zeta) \to \sum_{n=0}^{2n_\text{max}} \frac{f_n}{n!}\zeta^n\:.
\ee
This finite sum still gives a good approximation of $\cB f(\zeta)$ \textit{inside} the convergence circle.
To perform the Borel resummation, however, we have to integrate it along the whole positive real axis.
This means that we need the information on $\cB f(\zeta)$ \textit{outside} the convergence circle.
To resolve this problem, the Pad\'e approximant is usually used.
We replace the finite sum of the Borel sum by its ``diagonal'' Pad\'e approximant%
\footnote{Of course, one can also consider the ``non-diagonal'' Pad\'e approximant of the form $P_l(\zeta)/Q_m(\zeta)$ ($l\ne m$).
However, experience tells us that the diagonal Pad\'e approximant is usually the best one.}
\be
\sum_{n=0}^{2n_\text{max}} \frac{f_n}{n!}\zeta^n \to \frac{P_{n_\text{max}}(\zeta)}{Q_{n_\text{max}}(\zeta)} \:.
\ee
where $P_{n_\text{max}}(\zeta)$ and $Q_{n_\text{max}}(\zeta)$ are degree-$n_\text{max}$ polynomials.
Then, we can extrapolate the Pad\'e approximant outside the convergence circle.
The Pad\'e approximant also tells us the (approximate) singularity structure of the Borel transform. 
This numerically powerful procedure is often called the Borel-Pad\'e resummation.

\subsection{Figure-eight knot complement}
Let us start with the case of the figure-eight knot complement.
Note that this case has been studied in \cite{Gukov:2016njj} briefly,
but we find that there are a few small mistakes in their analysis.
We re-analyze it here in much more detail.
As a consequence, we arrive at a different conclusion from theirs.

We want to perform the Borel(-Pad\'e) resummation for the perturbative expansion
\eqref{eq:pert-fig8}.
As we will see just below, the perturbative expansion $Z_\text{pert}^\text{conj}(k)$
turns out to be Borel summable, and 
we find that its Borel resummation recovers the $S$-duality
for $k \leftrightarrow 1/k$.
On the other hand, the Borel resummation of $Z_\text{pert}^\text{geom}(k)$
does not.

\paragraph{Resumming the perturbative series.}
 
As in \eqref{eq:pert-fig8}, the perturbative expansions in the state-integral \eqref{state-integral-41} at $u=0$ are given by
\be
\ba
Z^{\rm geom}_{\rm  pert} (k)
= \frac{e^{\frac{k}{2\pi} V}}{3^{1/4}}\sum_{n=0}^\infty \frac{a_n^\text{geom}}{k^n}\:, \qquad
Z^{\rm conj}_{\rm  pert} (k)
= \frac{e^{-\frac{k}{2\pi} V}}{3^{1/4}}\sum_{n=0}^\infty \frac{a_n^\text{conj}}{k^n}\:,
\ea
\label{eq:pert-fig8-2}
\ee
where $V=\textrm{vol}(S^3\backslash \mathbf{4_1})=2 \im [\textrm{Li}_2 (e^{\pi i/3})]$.
Since the all the coefficients $a_n^\geom$ and $a_n^\conj$ are real,
we have the very simple relation (recall \eqref{eq:geom-conj})
\be
a_n^\text{conj}=(-1)^n a_n^\text{geom}\:.
\label{eq:geom-conj-2}
\ee
In spite of this simple relation, their resummations have quite different properties.

Following the method in \cite{Gukov:2016njj}, we computed the exact values of $a_n^\text{conj}$ up to $n=240$.
The first observation is that $Z^{\rm conj}_{\rm  pert}(k)$ is an alternating sum, while
$Z^{\rm geom}_{\rm  pert}(k)$ is a non-alternating one.
This implies that $Z^{\rm conj}_{\rm  pert}(k)$ is Borel summable, while
$Z^{\rm geom}_{\rm  pert}(k)$ is not.
To check this in detail, we analyze the singularities for the Borel-Pad\'e transform%
\footnote{Here the Borel transforms $\cB Z_\text{pert}^\text{geom, conj}(\zeta)$ are defined by
\[
\cB Z_\text{pert}^{\alpha}(\zeta)
=\sum_{n=0}^\infty \frac{a_n^{\alpha}}{n!}\zeta^n,\qquad
\alpha=\text{geom or conj}\:. 
\]} $\cB Z^{\rm conj}_{\rm  pert}(\zeta) \approx P_{n_\text{max}}(\zeta)/Q_{n_\text{max}}(\zeta)$.
In figure~\ref{fig:sing-1}, we show the pole structure of the denominator $Q_{n_\text{max}}(\zeta)$ of the Pad\'e approximant
for $n_\text{max}=100$ and $n_\text{max}=120$.
\begin{figure}[t]
\begin{center}
  \begin{minipage}[b]{0.45\linewidth}
    \centering
    \includegraphics[width=0.9\linewidth]{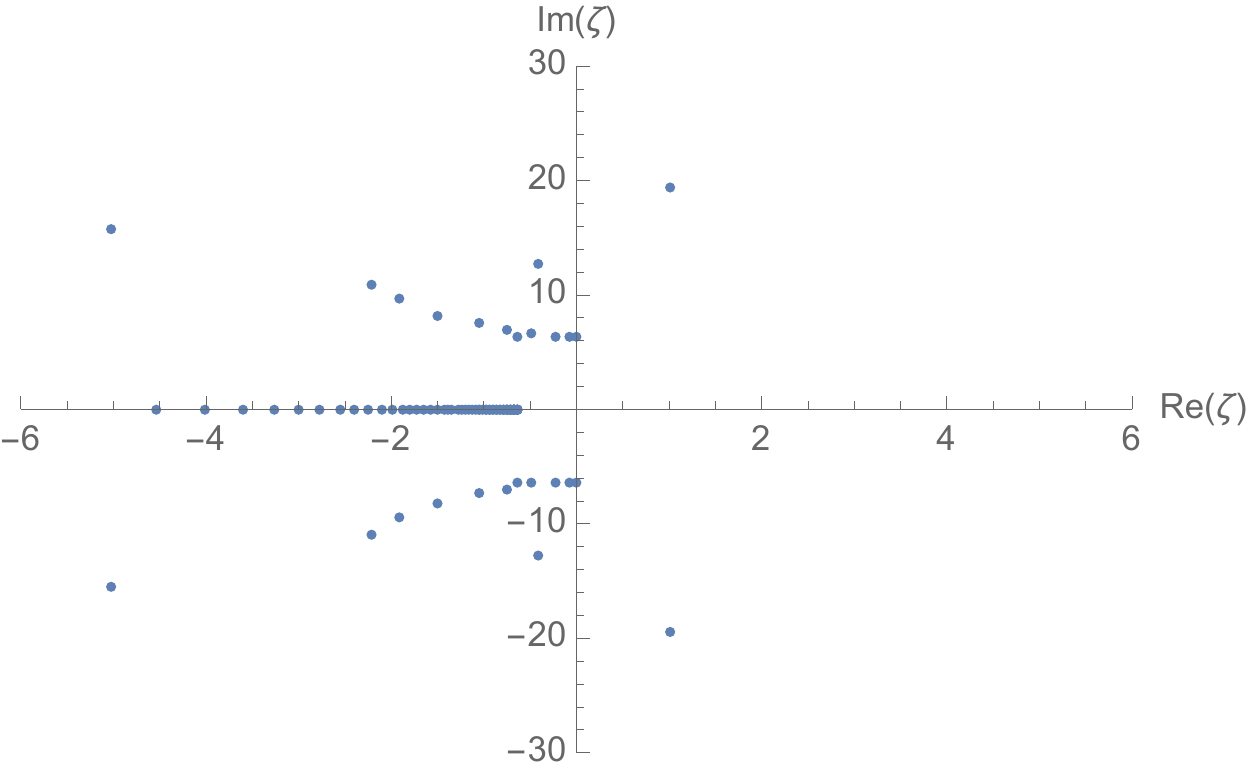}
  \end{minipage} \hspace{1cm}
  \begin{minipage}[b]{0.45\linewidth}
    \centering
    \includegraphics[width=0.9\linewidth]{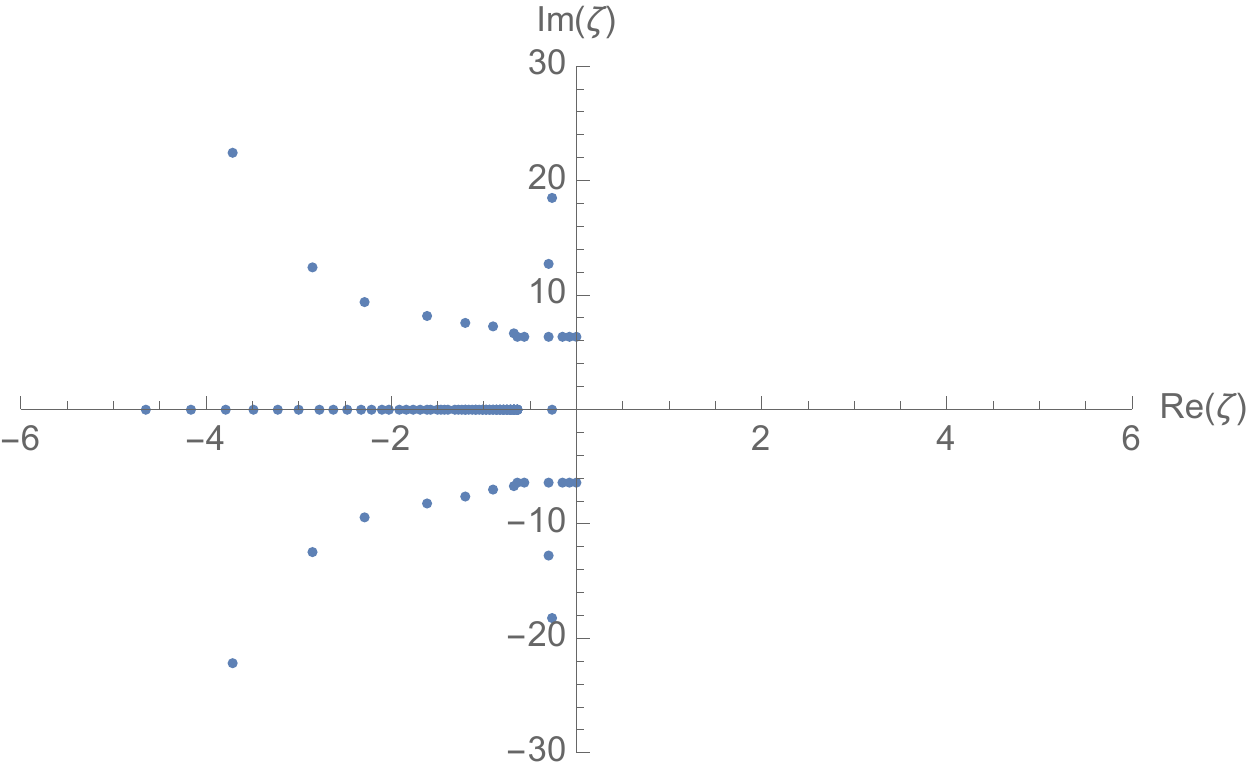}
  \end{minipage} 
\end{center}
  \caption{The pole structure of $Q_{n_\text{max}}(\zeta)$ for $n_\text{max}=100$ (left)
  and for $n_\text{max}=120$ (right) in the figure-eight knot complements.
  Some of poles depend on $n_\text{max}$, and they should not be the true singularities of 
  $\cB Z_\text{pert}^\text{conj}(\zeta)$. In the current case, we can conclude that $\cB Z_\text{pert}^\text{conj}(\zeta)$ does not have any singularities on the positive real axis but has on the negative real axis and on the imaginary axis.}
  \label{fig:sing-1}
\end{figure}
These figures strongly suggest that the Borel transform $\cB Z_\text{pert}^\text{conj}(\zeta)$
has no singularities on the positive real axis.
Using the relation \eqref{eq:geom-conj-2}, one easily finds the relation
\be
\cB Z_\text{pert}^\text{geom}(\zeta)=\cB Z_\text{pert}^\text{conj}(-\zeta)\:.
\ee
Since $\cB Z_\text{pert}^\text{conj}(\zeta)$ has singularities
on the negative real axis,%
\footnote{$\cB Z_\text{pert}^\text{conj}(\zeta)$ also seems to have singularities at $\zeta=\pm 2\pi i$.
These are not important in our analysis.} 
we conclude that $Z_\text{pert}^\text{geom}(k)$ is not Borel summable.

Let us proceed to the Borel resummation. What we actually do is the Borel-Pad\'e resummation
for $2n_\text{max}=240$:
\be
\cS Z_\text{pert}^\text{conj}(k)\approx \frac{e^{-\frac{k}{2\pi} V}}{3^{1/4}}
k \int_0^\infty d\zeta\, e^{-k\zeta} \frac{P_{120}(\zeta)}{Q_{120}(\zeta)}
\ee
For a given value of $k$, we can evaluate the Borel-Pad\'e resummation by this equation.
For example, the value at $k=1$ reads
\be
\cS Z_\text{pert}^\text{conj}(k=1) \approx 0.379567579522536528565367\dots.
\ee
We compare this value with
the direct evaluation of the state-integral \eqref{state-integral-41} along the contour in \eqref{Gamma41}.
For $u=0$ and $k=1$, we can deform the integration contour to the real axis, 
and the exact value of the state-integral was evaluated in \cite{Garoufalidis:2014ifa}
\be
\ba
Z(k=1;M=S^3\backslash \mathbf{4_1})(u=0) &=\frac{1}{\sqrt{3}} \left( e^{\frac{V}{2\pi}} -e^{-\frac{V}{2\pi}} \right) \\ 
&=0.3795675795225365285665625\dots .
\ea
\ee
We find agreement with 22-digit accuracy.%
\footnote{In \cite{Gukov:2016njj}, the authors conclude that the Borel resummation of $Z_\text{pert}^\text{conj}(k)$
does not reproduce the exact value of the state-integral. This conflicts our conclusion here.
The discrepancy comes from the exponential factor in \eqref{eq:pert-fig8-2}.
In \cite{Gukov:2016njj}, the exponential factor in $Z_\text{pert}^\text{conj}(k)$
is $e^{\frac{kV}{2\pi}}$.
It is however obvious that the exponential factor in $Z_\text{pert}^\text{conj}(k)$
must be $e^{-\frac{kV}{2\pi}}$ because $Z_\text{pert}^\text{conj}(k)$ is exponentially small 
in the semiclassical limit $k \to \infty$.
This factor is crucially important to reproduce the exact result for finite $k$
as well as the $S$-duality restoration below.}

More interestingly, we observe that the Borel resummation $\cS Z_\text{pert}^\text{conj}(k)$
has the $S$-duality relation:
\be
\cS Z_\text{pert}^\text{conj}(k)=\cS Z_\text{pert}^\text{conj}(1/k).
\ee
In fact, we show explicit values of $\cS Z_\text{pert}^\text{conj}(k)$
and $\cS Z_\text{pert}^\text{conj}(1/k)$ for various $k$'s in table~\ref{tab:Borel-resum-1}.
We also confirmed that all these values are in good agreement with the direct evaluation
of the state-integral \eqref{state-integral-41} for the contour \eqref{Gamma41}.

The integrand of the original state-integral \eqref{state-integral-41} possesses this symmetry manifestly,
but the perturbative expansion in $k \to \infty$, of course, 
makes this symmetry invisible. After the Borel resummation, the symmetry is precisely restored! 
We emphasize that to perform the Borel resummation, we use only the perturbative
data in $k \to \infty$.
Nevertheless the resummation ``knows'' the information in the opposite regime
$k \to 0$. 
This fact is surprising and unexpected. In fact, the authors in \cite{Gukov:2016njj}
did not expect this property.


%
%
%
\begin{table}[tb]
\caption{The $S$-duality restoration for the Borel-Pad\'e resummation of $Z_\text{pert}^\text{conj}(k)$.}
\begin{center}
\begin{tabular}{cll} \hline
$k$ & \hspace{1.5cm}$\cS Z_\text{pert}^\text{conj}(k)$ 
& \hspace{1.5cm}$\cS Z_\text{pert}^\text{conj}(1/k)$ \\
\hline
$\sqrt{2}$ & $\textbf{0.36542977253384313}898$ & $\textbf{0.36542977253384313}647$ \\
$\sqrt{3}$ & $\textbf{0.3445028183404}9000808$ & $\textbf{0.3445028183404}8996022$ \\
$2$ & $\textbf{0.32447273598566}357884$ & $\textbf{0.32447273598566}448145$ \\
$\sqrt{5}$ & $\textbf{0.3062748854294}4963878$ & $\textbf{0.3062748854294}6706198$ \\
$\sqrt{6}$ & $\textbf{0.289874501536}33354513$ & $\textbf{0.289874501536}46160882$ \\
\hline
\end{tabular}
\end{center}
\label{tab:Borel-resum-1}
\end{table}%

Next, let us discuss the Borel resummation of $Z_\text{pert}^\text{geom}(k)$.
As we have already seen, $Z_\text{pert}^\text{geom}(k)$ is not Borel summable.
Therefore we have to consider the deformed Borel resummations \eqref{eq:lateral-Borel}.
In the actual computation, we use the Borel-Pad\'e resummations:
\be
\cS_{\pm} Z_\text{pert}^\text{geom}(k)\approx \frac{e^{\frac{k}{2\pi} V}}{3^{1/4}}
k \int_0^{\infty e^{\pm i \epsilon}} d\zeta\, e^{-k\zeta} \frac{P_{120}(-\zeta)}{Q_{120}(-\zeta)}
\ee
where the Pad\'e approximant is the same function appearing in $Z_\text{pert}^\text{conj}(k)$.
These Borel resummations turn out to be complex-valued.
For example, the values at $k=1$ are given by
\be
\cS_\pm Z_\text{pert}^\text{geom}(1)\approx 1.0526393020 \pm 0.5693505539i\:.
\ee
Moreover, we observe that the Borel resummations $\cS_{\pm} Z_\text{pert}^\text{geom}(k)$
do not have the S-dual symmetry:
\be
\cS_{\pm} Z_\text{pert}^\text{geom}(k) \ne \cS_{\pm} Z_\text{pert}^\text{geom}(1/k)\:.
\ee
For instance, for $k=\sqrt{2}$, we have
\be
\ba
\cS_{\pm} Z_\text{pert}^\text{geom}(\sqrt{2})&\approx 1.358610063 \pm 0.548144707 i \:, \\
\cS_{\pm} Z_\text{pert}^\text{geom}(1/\sqrt{2})&\approx 0.7999826621 \pm 0.5481068265 i \:.
\ea
\ee
\begin{figure}[t]
\begin{center}
  \begin{minipage}[b]{0.45\linewidth}
    \centering
    \includegraphics[width=0.9\linewidth]{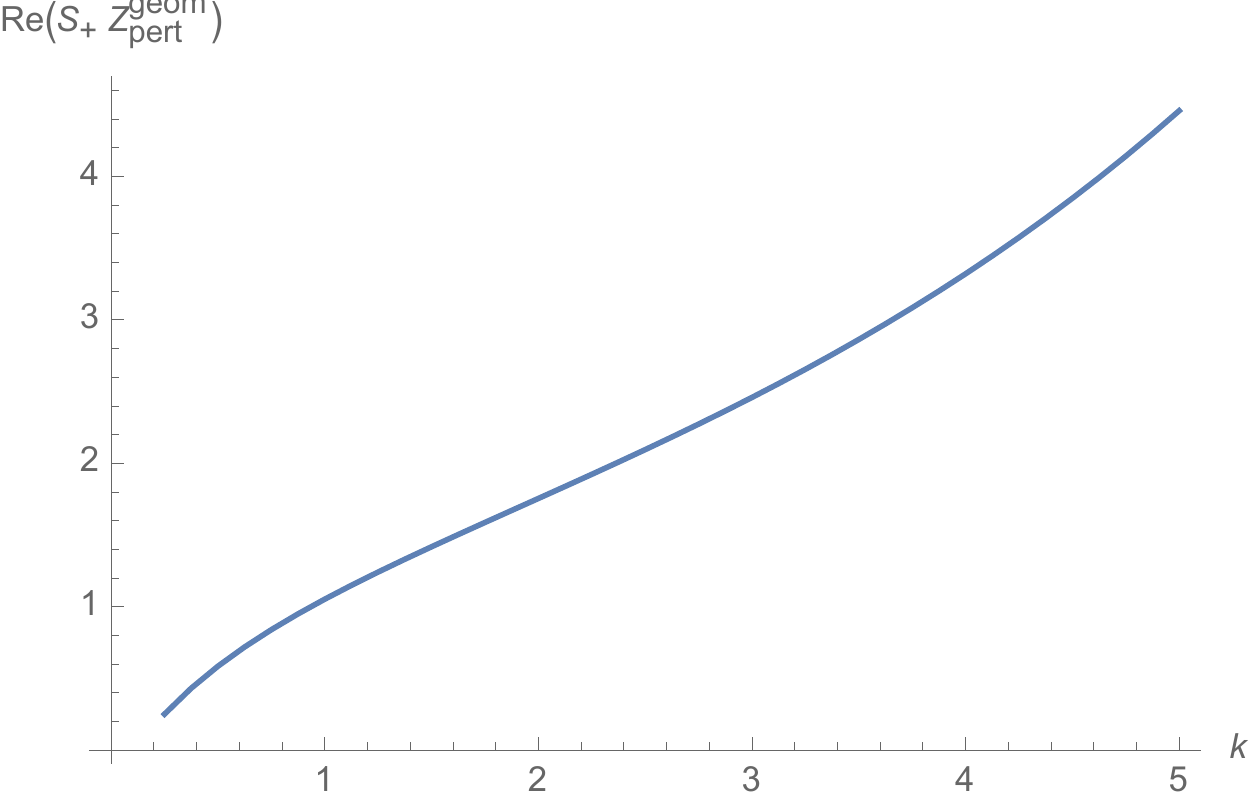}
  \end{minipage} \hspace{1cm}
  \begin{minipage}[b]{0.45\linewidth}
    \centering
    \includegraphics[width=0.9\linewidth]{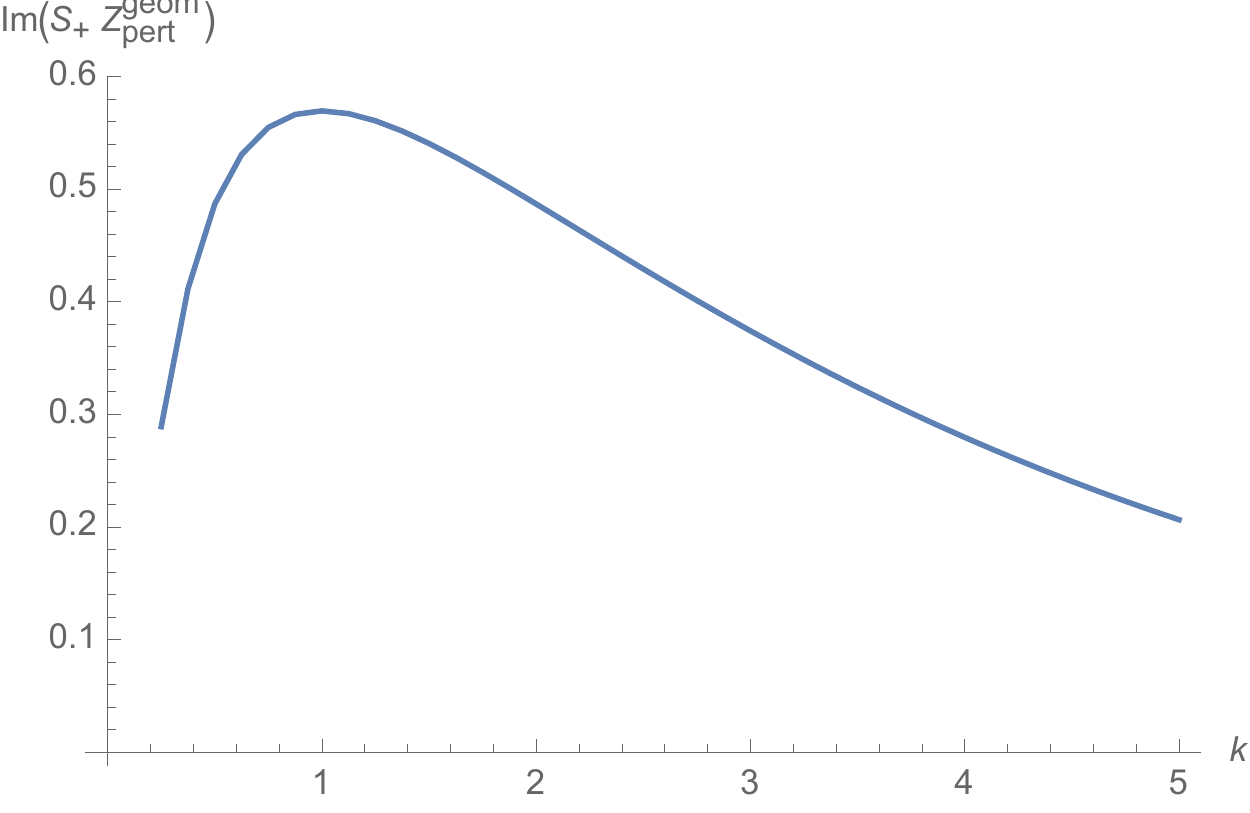}
  \end{minipage} 
\end{center}
  \caption{The real part of $\cS_+ Z_\pert^\geom(k)$ does not recover the $S$-duality, while the imaginary part does.}
  \label{fig:k-dep}
\end{figure}
We show the $k$-dependence of the real and imaginary parts of $\cS_+ Z_\pert^\geom(k)$ in figure~\ref{fig:k-dep}.
Though $\cS_{\pm} Z_\text{pert}^\text{geom}(k)$
do not have the S-dual relation totally, their imaginary part seems to have it.
This is because the imaginary part is precisely related to the Borel sum $\cS Z_\text{pert}^\text{conj}(k)$.
In fact, the standard resurgent analysis (see \cite{Dorigoni:2014hea} for instance) tells us that the difference
of $\cS_{\pm} Z_\text{pert}^\text{geom}(k)$ is given by
\be
\cS_{+} Z_\text{pert}^\text{geom}(k)-\cS_{-} Z_\text{pert}^\text{geom}(k)
=S \cdot \cS Z_\text{pert}^\text{conj}(k)\:,
\label{eq:discon}
\ee
where $S$ is called a Stokes constant.
As we will see below, in our case we have $S=3i$.

\paragraph{Large order behavior.}
Finally, we discuss the large order behavior of the perturbative expansion.
From the resurgent analysis, the large order behavior of $Z_\text{pert}^\text{conj}(k)$
provides the information on the other saddle $Z_\text{pert}^\text{geom}(k)$.
More precisely, as in \cite{Gukov:2016njj}, we expect the large order behavior
\be
a_n^\text{conj}=\frac{S}{2\pi i} \frac{(n-1)!}{A^n} \left[ 1+\frac{a_1^\text{geom} A}{n-1}
+\frac{a_2^\text{geom} A^2}{(n-1)(n-2)}+\cdots \right], \qquad
n \to \infty \:.
\label{eq:LO}
\ee
Since we have $a_n^\text{conj}$ up to $n=240$,
we can extract the information on $A$, $S$ and $a_n^\text{geom}$
very precisely from this formula.

To know $A$, we look at a relation
\be
\frac{n a_n^\text{conj}}{a_{n+1}^\text{conj}}=A+\cO(n^{-2}), \qquad n \to \infty \:.
\ee
To accelerate the convergence of this sequence,
we use the Richardson extrapolation.
For the analysis of the large order behavior
by using the Richardson extrapolation, see \cite{Marino:2007te}.
Let us define the $m$-th Richardson transform of a given sequence $f_n$ by
\be
\cR_m [f_n]:=\sum_{k=0}^m \frac{ (-1)^{k+m}(n+k)^m}{k! (m-k)!}f_{n+k}.
\ee
If the sequence $f_n$ behaves as
\be
f_n=C [1+\cO(n^{-1}) ], \qquad n \to \infty,
\ee
then the Richardson transform of $f_n$ behaves as
\be
\cR_m [f_n]=C [1+\cO(n^{-m-1}) ], \qquad n \to \infty.
\ee
Therefore the convergence speed is improved.

In the current case, we apply the 80th Richardson transform%
\footnote{To compute the $m$-th Richardson transform of $f_n$, we need the higher elements 
$f_{n+1},\dots, f_{n+m}$.
If we have $f_n$ up to $n=N_\text{max}$,
we can perform the $m$-th Richardson transform up to $n'=N_\text{max}-m$.
We choose $m$ as good convergence as possible.}
to the sequence $n a_n^\text{conj}/a_{n+1}^\text{conj}$, and
find the convergent value
\be
\cR_{80} [159 a_{159}^\text{conj}/a_{160}^\text{conj}]  = -0.646131894438901\dots.
\ee
As found in \cite{Gukov:2016njj}, the exact value of $A$ is given by the difference of the actions of the two saddles
$A^\text{geom}$ and $A^\text{conj}$,
\be
A=-\frac{V}{\pi}=-0.646131894438901\dots.
\ee
We find remarkable agreement with $|A-\cR_{80} [159 a_{159}^\text{conj}/a_{160}^\text{conj}] |
\sim \cO(10^{-98})$.
This $A$ is also related to a singularity on the Borel transform $\cB Z_\text{pert}^\text{conj}(\zeta)$.
The closest singularity of the Pad\'e approximant of $\cB Z_\text{pert}^\text{conj}(\zeta)$ 
on the negative real axis from the origin%
\footnote{In the right of figure~\ref{fig:sing-1}, one can see a pole on the negative real axis
at $\zeta \approx -0.27$.
This pole however does not appear in the left figure, and is considered to be a ``false'' singularity.}
is
\be
\zeta \approx -0.6462,
\ee
which is indeed in agreement with $A$.

Once the exact value of $A$ is known, we can extract $S$ by
\be
b_n :=2\pi \frac{A^n a_n^\text{conj}}{(n-1)!}= S/i +\cO(n^{-1}), \qquad n \to \infty \:.
\ee
Using the Richardson transform of $b_n$ again, we find
\be
\cR_{80}[b_{160}]=3+ \cO(10^{-97}).
\ee
This strongly suggest that the exact value of $S$ is
\be
S=3i.
\ee
Note that our obtained value is different from the one in \cite{Gukov:2016njj}.
The value in \cite{Gukov:2016njj} is $S^\text{[GMP]} \approx 7.51989 i$.%
\footnote{This numerical value is very likely $S^\text{[GMP]}=3\sqrt{2\pi} i$.
The factor $\sqrt{2\pi}$ comes from the Gaussian integral normalization. We thank M. Mari\~no for this point.}
The evidence of our result here is that the discontinuity \eqref{eq:discon} holds only for $S=3i$.

Repeating this way, one can confirm the large order relation \eqref{eq:LO} with very high numerical accuracy.

\subsection{Thurston manifold}

\paragraph{Borel resummation.}
In this case, the perturbative expansions of the state-integral \eqref{state-integral-thurston1} (or \eqref{state-integral-thurston2}) take the forms
\be
\ba
\Zpc(k)&=e^{-\frac{k\cA}{2\pi}} \cN \sum_{n=0}^\infty \frac{a_n^\text{conj}}{k^n} \:,\\
\Zpg(k)&=e^{\frac{k\cA^*}{2\pi}} \cN^* \sum_{n=0}^\infty \frac{a_n^\text{geom}}{k^n} \:,
\ea
\ee
where $a_0^\conj=a_0^\geom=1$ and
\be
\cA \approx 0.981369 - 1.52067 i, \qquad
\cN \approx -0.102535 + 0.701692 i.
\ee
Recall that we have the relation \eqref{eq:geom-conj}.
Also, we can compute the perturbative expansion of the state-integral \eqref{state-integral-thurston2}
around the saddle \eqref{eq:z-saddle-conj}.
As mentioned before, the result coincides with $\Zpc(k)$:
\be
\widetilde{Z}_\pert(k)=e^{-\frac{k\cA}{2\pi}}\cN \sum_{n=0}^\infty \frac{a_n^\conj}{k^n}\:,
\ee
In the following, we mainly focus on the resummation of $\widetilde{Z}_\pert(k)$.
From the state-integral expression \eqref{state-integral-thurston2},
we computed the numerical values of $a_n^\conj$ up to $n=100$.
Using these data, we show in figure~\ref{fig:sing-2} the singularities of the Pad\'e approximant of the Borel transform
\be
\cB \widetilde{Z}_\pert(\zeta)=\sum_{n=0}^\infty \frac{a_n^\text{conj}}{n!}\zeta^n \:.
\label{eq:Borel-Thurston}
\ee
It is very likely that $\widetilde{Z}_\pert(k)$ is Borel summable.
\begin{figure}[t]
\begin{center}
  \begin{minipage}[b]{0.45\linewidth}
    \centering
    \includegraphics[width=0.9\linewidth]{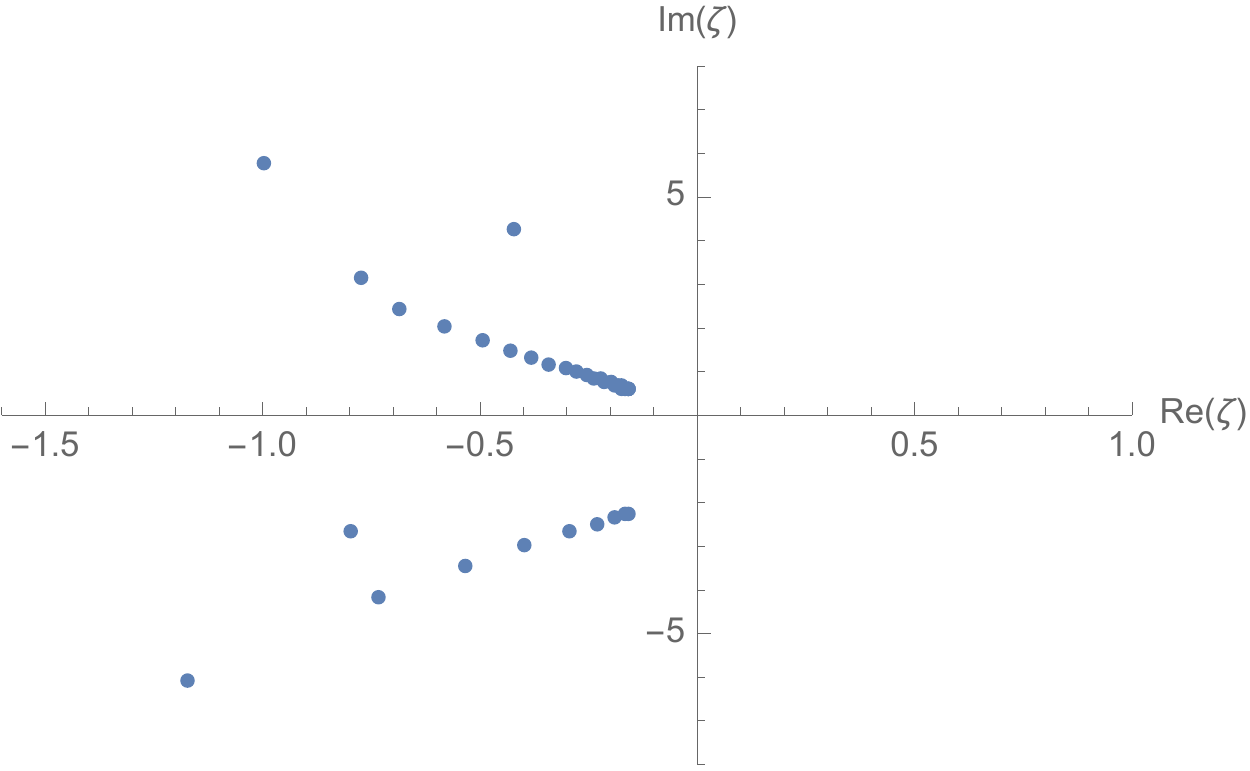}
  \end{minipage} \hspace{1cm}
  \begin{minipage}[b]{0.45\linewidth}
    \centering
    \includegraphics[width=0.9\linewidth]{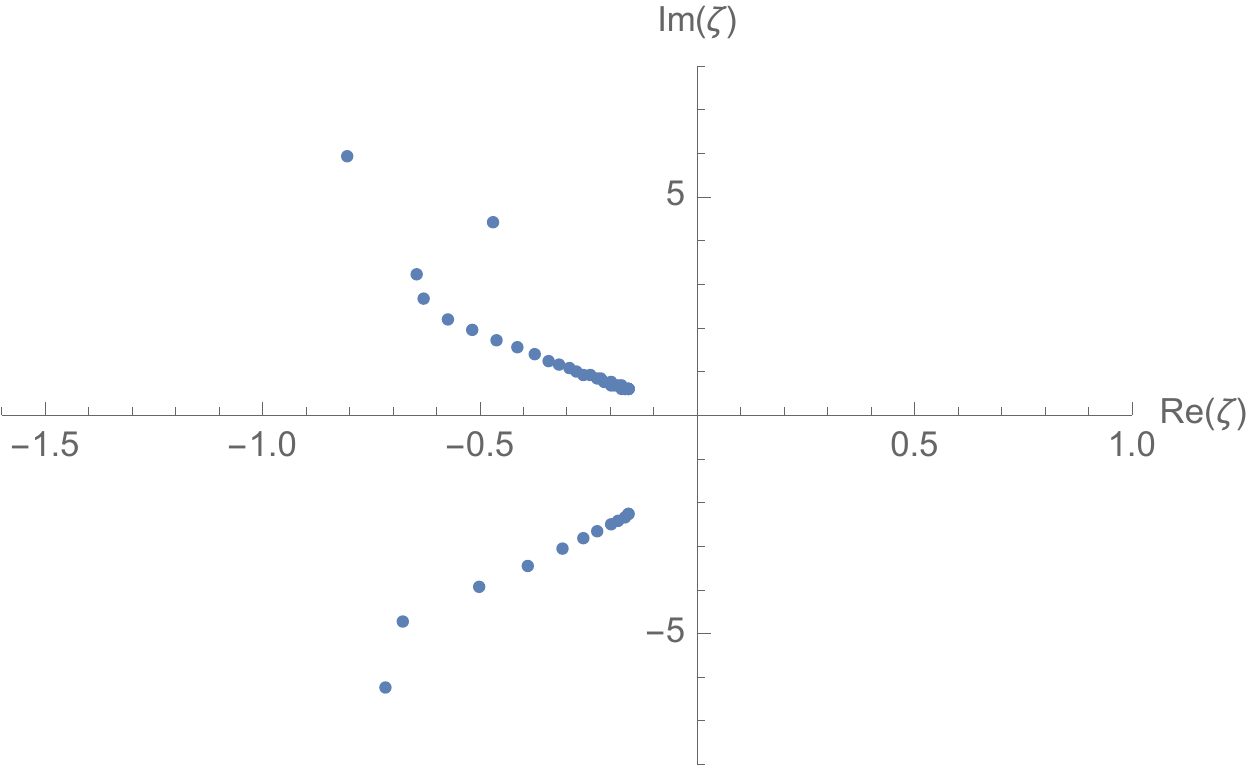}
  \end{minipage} 
\end{center}
  \caption{The pole structure of $Q_{n_\text{max}}(\zeta)$ in the Thurston manifold \eqref{eq:Borel-Thurston} for $n_\text{max}=40$ (left)
  and for $n_\text{max}=50$ (right).}
  \label{fig:sing-2}
\end{figure}
It is observed that the closest singularity from the origin is located at
\be
\zeta \approx -0.1563 + 0.5846 i .
\label{eq:closest-sing}
\ee
The real part is in agreement with that of the action: $\real [\cA/(2\pi)]=0.1561897\dots$.

We perform the Borel-Pad\'e resummation%
\footnote{In the computation, we keep all the numerical values sufficiently high precision.} by
\be
\cS \widetilde{Z}_\pert(k) \approx e^{-\frac{k\cA}{2\pi}}\cN\,
k \int_0^\infty d\zeta \, e^{-k \zeta} \frac{P_{50}(\zeta)}{Q_{50}(\zeta)}\:.
\ee
For $k=1$, we get
\be
\cS \widetilde{Z}_\pert(1) \approx -0.2898929700 + 0.4325462447 i .
\ee
This result is compared with the numerical evaluation of the state-integral \eqref{state-integral-thurston2}.
%
%
We evaluate it along $\Gamma_{\rm Thurston}$ in \eqref{Gamma-thurston}.
Then, we find
\begin{align}
\widetilde{Z} (1)\approx  -0.2898929693 + 0.4325462442 i .
\end{align}
We also confirm the $S$-duality restoration in the Borel resummation, as shown in table~\ref{tab:Borel-resum-2}.
\begin{table}[tb]
\caption{The $S$-duality restoration for the Thurston manifold.}
\begin{center}
\begin{tabular}{cll} \hline
$k$ & \hspace{1.5cm}$\cS \widetilde{Z}_\pert(k)$ 
& \hspace{1.5cm}$\cS \widetilde{Z}_\pert(1/k)$ \\
\hline
$\sqrt{2}$ & $-{\bf 0.2972319}065 + {\bf 0.4156327}694 i$ & $-{\bf 0.2972319}477 + {\bf 0.4156327}825 i$ \\
$\sqrt{3}$ & $-{\bf 0.307297}4784 + {\bf 0.3896472}188 i$ & $-{\bf 0.307297}8009 + {\bf 0.3896472}596 i$ \\
$2$          &  $-{\bf 0.31593}58614 + {\bf 0.3636643}427 i$ & $-{\bf 0.31593}70863 + {\bf 0.3636643}208 i$ \\
$\sqrt{5}$ & $-{\bf 0.3228}283893 + {\bf 0.339069}5105 i$ & $-{\bf 0.3228}315994 + {\bf 0.339069}0679 i$ \\
$\sqrt{6}$ & $-{\bf 0.32816}14035 + {\bf 0.3160}611133 i$ & $-{\bf 0.32816}81342 + {\bf 0.3160}594714 i$ \\
\hline
\end{tabular}
\end{center}
\label{tab:Borel-resum-2}
\end{table}%

In the case of the Thurston manifold, the perturbative expansion $\Zpg(k)$ is also Borel summable.
We again observe that the Borel resummation of $\Zpg(k)$ does not reproduce the $S$-dual relation.
For $k=\sqrt{2}$, we have
\be
\ba
\cS \Zpg(\sqrt{2}) &\approx 0.0932017344 - 0.8386515714 i, \\
\cS \Zpg(1/\sqrt{2}) &\approx -0.1378028444 - 0.6725116917 i.
\ea
\ee

\paragraph{Large order behavior.}
Let us proceed to the large order behavior.
As in \eqref{eq:LO}, we assume the large order behavior of the form
\be
a_n^\conj=\frac{S}{2\pi i} \frac{(n-1)!}{A^n} \left[ 1+\frac{b_1 A}{n-1}
+\frac{b_2 A^2}{(n-1)(n-2)}+\cdots \right], \qquad  n \to \infty.
\ee
Using the first 100 coefficients, we find the numerical values
\be
\ba
A &\approx -0.1561897001 + 0.5841922570 i, \\
S &\approx 0.1683579878 - 0.0246012135 i, \\
b_1 & \approx 0.6626248474 i,\qquad
b_2 \approx -0.8896638449.
\ea
\ee
All of these values are stable in the 17th Richardson transform at least up to this digit.
One can see that the value of $A$ coincides with the closest singularity \eqref{eq:closest-sing},
as in the figure-eight knot complement.
Our analysis implies that $b_1$ seems purely imaginary and that $b_2$ seems real.
So far, it is unclear to us the relation between this large order behavior and
the saddle-point approximation in the state-integral \eqref{state-integral-thurston2} 
(or \eqref{state-integral-thurston1}). This is not a main purpose in this paper.
It would be interesting to explore it in more detail.

\subsection{Physical reasoning of the $S$-duality}
In the previous subsection, we saw that the perturbative expansion around the saddle corresponding to the flat connection $A=A^\conj$
is Borel summable and that its Borel resummation has the $S$-duality.
We also observed that the perturbation around $A^\geom$ is not Borel summable for the figure eight knot complement,
but Borel summable for the Thurston manifold.

In general, it is not simple to say whether the perturbation around a given flat connection $A^\alpha$ is Borel summable or not.
Nevertheless, we can say that for the particular connection $A^\conj$, the perturbative expansion $Z_\pert^\conj$ is always Borel summable.
The reason is as follows.
If $Z_\pert^\conj$ is not Borel summable, then it has to receive non-perturbative corrections to cancel the ambiguity of the Borel sum.
However, the inequality \eqref{eq:ineq-CS} shows that there are no saddles, whose exponentiated classical actions 
are smaller than that for $A^\conj$:
\be
e^{-\frac{k}{4\pi} \im[CS[A^\conj;M]]}<e^{-\frac{k}{4\pi} \im[CS[A^\alpha;M]]}<e^{-\frac{k}{4\pi} \im[CS[A^\geom;M]]}.
\ee
This means that $Z_\pert^\conj$ does not receive any non-perturbative corrections, and we conclude that $Z_\pert^\conj$ must be Borel summable.
Our conjecture here is that the Borel resummation $\cS Z_\pert^\conj$ has the $S$-dual symmetric structure in $k\leftrightarrow 1/k$.
In this sense, the flat connection $A^\conj$ is very special.

The emergence of the $S$-duality after the Borel resummation of $Z^{\rm conj}_{\rm pert}$ is somewhat surprising since there is no such a symmetry in the  path integral \eqref{Formal path-integral}. One heuristic explanation  of this surprise is the following.  First, notice that the integrand in \eqref{Formal path-integral} is not unitary for complex gauge field $A$ and  the path-integral  makes sense only at the perturbative level. 
As will be explained below, there is  a unitary   $SL(2)$ CS theory whose partition function has the same asymptotic expansion as  \eqref{Formal path-integral} in a certain limit of coupling in the theory. So the unitary theory can be considered as a non-perturbative completion of the formal perturbative CS partition function in \eqref{Formal path-integral}.  Further, the unitary complex CS theory  has an $S$-duality as a manifest symmetry. The Borel resummation  somehow knows the non-perturbative completion and  it gives the non-perturbative answer which has the $S$-duality symmetry. 

Let us explain the unitary complex CS theory in more detail. The complex $SL(2)$ CS theory depends on two CS levels, $K$ and $\sigma$, whose action is given by %
\begin{align}  
\frac{K +  \sigma}{8\pi } CS[A;M] + \frac{K-  \sigma}{8\pi} CS[\tilde{A};M] \label{lagrangian-complex-CS}
\end{align}
For the invariance under the large gauge transformation, $K$ should be an integer:
\begin{align}
K\in \mathbb{Z}\;.
\end{align}
For the unitarity of the theory, $\sigma$ is either real or purely imaginary. 
\begin{align}
\sigma \in \mathbb{R}\quad \textrm{or}\quad   \sigma \in i\mathbb{R}\;.
\end{align}
In \cite{Dimofte:2015kkp}, it is conjectured that if we choose
\begin{align}
K=1\;,\qquad \sigma = \frac{1-b^2}{1+b^2}\;, \label{K=1 CS}
\end{align}
then the asymptotic expansion of the partition function of the complex $SL(2)$ CS  theory in a singular limit $b\rightarrow 0$ is equivalent to the formal perturbative expansion in \eqref{Formal path-integral} with identification $k=b^{-2}$. So, the  $SL(2)$ CS theory with $K=1$ can be considered as a non-perturbative completion of the formal path integral in \eqref{Formal path-integral}. 
After the substitution in \eqref{K=1 CS}, the action \eqref{lagrangian-complex-CS}  becomes
\begin{align}
\frac{1}{4\pi (1+b^2)} CS[A;M]+ \frac{1}{4\pi (1+b^{-2})} CS[\tilde{A};M]\;, \label{Action for K=1}
\end{align}
which has the following $S$-duality,
\begin{align}
(b, A) \; \leftrightarrow (b^{-1},\tilde{A})\;.
\end{align}
In the limit $b\rightarrow 0$, the second term in \eqref{Action for K=1} vanishes and the action reduced to a CS action only with $A$ with a quantum parameter $\hbar = 2\pi i (1+b^2)$. The resulting action is equivalent to the action in \eqref{Formal path-integral} except for $k=b^{-2}$ is replaced by $(1+b^2)^{-1}$. In a quantization of CS theory, the relevant quantum parameter is $q:=e^{\hbar}$  instead of $\hbar$ and the  difference between two actions  disappears. This is an heuristic derivation of the conjecture in \cite{Dimofte:2015kkp}. 

\acknowledgments
The authors would like to thank S.~Gukov and P.~Putrov for discussions. We also thank M.~Mari\~no for reading the manuscript. The contents of this paper were presented by one (DG) of authors in ``The International Workshop on Superconformal Theories 2017 (SCFT2017)'' at Sichuan University and  we thank the audience for feedback. The work of DG was supported by Samsung Science and
Technology Foundation under Project Number SSTBA1402-08. 
The work of YH is supported by Rikkyo University Special Fund for Research.

\appendix

\section{Quantum dilogarithms}\label{app:QDL}
In this appendix, we briefly summarize some basic properties of quantum dilogarithms that
we need in the main text.

\paragraph{Compact quantum dilogarithm.}
First, we define the compact quantum dilogarithm by
\be
\phi_q(X):=(X;q)_{\infty}=\prod_{n=0}^\infty (1-Xq^n)
=\exp \biggl[ -\sum_{k=1}^\infty \frac{X^k}{k(1-q^k)} \biggr]\qquad
(|q|<1)\:.
\ee
For $q=e^{\hbar}$, this function has the following semiclassical expansion:
\be
\log \phi_q(X)=\sum_{n=0}^\infty
\frac{B_{n}}{n!} \Li_{2-n}(X) \hbar^{n-1} \:,
\ee
where $B_{n}$ is the $n$-th Bernoulli number.
Of course, in the classical limit $\hbar \to 0$, $\log \phi_q(X)$
reduces to the classical dilogarithm.

\paragraph{Non-compact quantum dilogarithm.}
We also define the non-compact (or Faddeev's) quantum dilogarithm by
\be
\Phi_b(z):= \exp \biggl[
\int_{\mathbb{R}+i\epsilon} \frac{dt}{t} \frac{e^{-2itz}}{4\sinh(bt) \sinh(b^{-1} t)} \biggr]\:.
\label{eq:NQDL}
\ee
By this definition, it is obvious to see that the function has an important symmetry
\be
\Phi_b(z)=\Phi_{b^{-1}}(z)\:.
\ee
For $b=1$, the function reduces to the classical (di)logarithm:
\be
\Phi_{b=1}(z)=\exp \biggl[ iz \log(1-e^{2\pi z})+\frac{i}{2\pi} \Li_2(e^{2\pi z}) \biggr] \:.
\ee
Note that compared to the compact quantum dilogarithm,
the non-compact one is well-defined even for $|q|=1$.
For $|q|<1$, it is constructed by two copies of
the compact quantum dilogarithm:
\be
\Phi_b(z)=\frac{\phi_q(-q^{1/2}e^{2\pi b z})}{\phi_{\tilde{q}^{-1}}(-\tilde{q}^{-1/2}e^{2\pi z/b})}
=\exp \biggl[ \sum_{k=1}^\infty \frac{(-1)^k e^{2\pi k bz}}{k(q^{k/2}-q^{-k/2})}
+\sum_{k=1}^\infty \frac{(-1)^k e^{2\pi k z/b}}{k(\tilde{q}^{k/2}-\tilde{q}^{-k/2})} \biggr] \:,
\label{eq:NC-C}
\ee
where
\be
q:=e^{2\pi i b^2},\qquad \tilde{q} := e^{2\pi i/b^2} \:.
\ee
As in the compact case, we can expand $\Phi_b(z)$ around $\hbar=2\pi i b^2=0$,
\be
\log \Phi_b(z)=\sum_{n=0}^\infty \frac{B_{2n}(1/2)}{(2n)!}\Li_{2-2n}(-e^{2\pi b z} ) \hbar^{2n-1},
\qquad \hbar \to 0,
\label{eq:sc-1}
\ee
where $B_n(x)$ is the Bernoulli polynomial.
Note that the compact function $\phi_q(-q^{1/2} e^{2\pi b z})$ also has the same semiclassical expansion:
\be
\log \phi_q(-q^{1/2} e^{2\pi b z})=\sum_{n=0}^\infty \frac{B_{2n}(1/2)}{(2n)!}\Li_{2-2n}(-e^{2\pi b z} ) \hbar^{2n-1},
\qquad  \hbar \to 0.
\label{eq:sc-2}
\ee
This is a consequence of the relation \eqref{eq:NC-C}.
At the semiclassical level, we cannot distinguish the non-compact function $\Phi_b(z)$
from the compact one $\phi_q(-q^{1/2} e^{2\pi b z})$.
However, one should keep in mind that the equations \eqref{eq:sc-1} and \eqref{eq:sc-2} mean
the equalities in the asymptotic sense in $\hbar \to 0$.
We know, of course, $\Phi_b(z) \ne \phi_q(-q^{1/2} e^{2\pi b z})$ for finite $\hbar$.

\paragraph{Resummation.}
As shown in \cite{Hatsuda:2015owa}, the semiclassical expansion on the right hand side in \eqref{eq:sc-1} or \eqref{eq:sc-2}
is resummed exactly.
The resummed function turns out to reproduce the \textit{non-compact} quantum dilogarithm, not the compact one.
Let us denote the semiclassical expansion as
\be
L_b(z):=\sum_{n=0}^\infty \frac{B_{2n}(1/2)}{(2n)!}\Li_{2-2n}(-e^{2\pi b z} ) (2\pi i b^2)^{2n-1}\:.
\label{eq:sc-3}
\ee
The basic idea for the resummation is to use the following identity
\be
B_{2n}(1/2)
=(-1)^n 4n \int_0^\infty dx \frac{x^{2n-1}}{e^{2\pi x}+1} ,\qquad
n \geq 1.
\ee
Plugging it into \eqref{eq:sc-3} and exchanging the sum and the integral, we get
\be
L_b^\text{resum}(z)=\frac{\Li_2(-e^{2\pi b z})}{2\pi i b^2}
-i\int_{0}^\infty \frac{dx}{e^{2\pi x}+1} \log \left( \frac{1+e^{2\pi b z-2\pi b^2 x}}{1+e^{2\pi b z+2\pi b^2 x}} \right) \:.
\label{eq:Lb-resum-0}
\ee 
This is further rewritten as a simpler form
\be
L_b^\text{resum}(z)=\frac{i}{2\pi} \int_{-\infty}^\infty \frac{dt}{e^t+1} \log(1+e^{2\pi b z+b^2 t})\:.
\label{eq:Lb-resum}
\ee
Surprisingly, this resummation recovers the symmetry for $b \leftrightarrow b^{-1}$:
\be
L_b^\text{resum}(z)=L_{b^{-1}}^\text{resum}(z).
\ee
We stress that this symmetry restoration is far from obvious in the integral representation \eqref{eq:Lb-resum-0} or \eqref{eq:Lb-resum}.
We have checked it numerically.
Finally, one can also numerically confirm that this resummation reproduces the original non-compact quantum dilogarithm:
\be
\log \Phi_b(z)=L_b^\text{resum}(z)=\frac{i}{2\pi} \int_{-\infty}^\infty \frac{dt}{e^t+1} \log(1+e^{2\pi b z+b^2 t}) .
\ee


\bibliographystyle{JHEP}
\bibliography{Draft_v1}



\end{document}